\documentclass[preprint,superscriptaddress,showpacs,preprintnumbers,amsmath,amssymb]{revtex4}
\usepackage[dvips]{graphicx}
\usepackage{epsfig}                
\usepackage{colordvi}              

\begin{document}

\title{Crystalline multilayers of the confined Yukawa system}

\author{Erdal Celal O\u guz, Ren\'e Messina, Hartmut L\"owen}
\affiliation{%
Institut f\"ur Theoretische Physik II: Weiche Materie,
Heinrich-Heine-Universit\"at
D\"usseldorf, Universit\"atsstra\ss e 1, D-40225 D\"usseldorf, Germany\\
}%

\date{\today}

\begin{abstract}
The phase diagram of Yukawa particles confined between two parallel hard walls
is calculated at zero-temperature beyond
the bilayer regime by lattice-sum-minimization. Tuning the 
screening, a rich phase behavior is found in the regime bounded by stable 
two-triangular layers and 3-square layers.
In this regime, alternating prism phases with square and triangular basis,
structures derived from a hcp bulk lattice, and a structure with two outer layers and 
two inner staggered rectangular layers, reminiscent of a Belgian waffle iron, are stable. 
These structures are verifiable in experiments on charged colloidal suspensions and dusty plasma sheets.
\end{abstract}


\maketitle


If a system is confined to a thin slit geometry, its properties  are 
drastically altered with respect to the bulk \cite{Haase,Klafter}. In particular, 
the freezing transition in confined geometry occurs into multilayered crystals 
whose structure is dictated by an interplay of the interparticle interaction and the confining potential 
\cite{Zangi,Ayappa,Cornelissens}. 
For {\it hard spheres} between two parallel hard plates of finite separation $D$, 
different solid lattices are getting stable 
with the following cascade for increasing width $D$: 
If $D$ equals the hard sphere diameter, a crystalline monolayer 
is stable which possesses 
a two-dimensional triangular ($\Delta$) lattice symmetry. This layer buckles 
\cite{Schmidt1,Nelson,Frey} upon increasing $h$ such that a two-layer
situation is realized. For higher $h$, there are stable  structures that
correspond to two intersecting square lattices ($2\square$) or triangular lattices 
($2\Delta$) \cite{Pieranski, Schmidt2} with a two-layer rhombic phase in between \cite{Schmidt1,Schmidt2}. 
Beyond bilayers, the transition from the $2\Delta$ to the $3\square$ (simple quadratic trilayer) phase are mediated by 
prism phases which are made up of alternating prism-like arrays \cite{Neser,Fortini}. In experiments on strongly screened colloidal suspensions, 
however, crystals derived from the hexagonal-closed-packed geometry were found as interpolating structures
which correspond to four or more layers \cite{Manzano_2007,Fontecha_2007}. Beyond the $3\square$ phase, 
the structures are even more complicated \cite{Fortini,Manzano_2006,Schope_2006,Fontecha_2007}.

\begin{figure*}[t]
  \begin{center}
    \includegraphics[width=13cm,height=12cm]{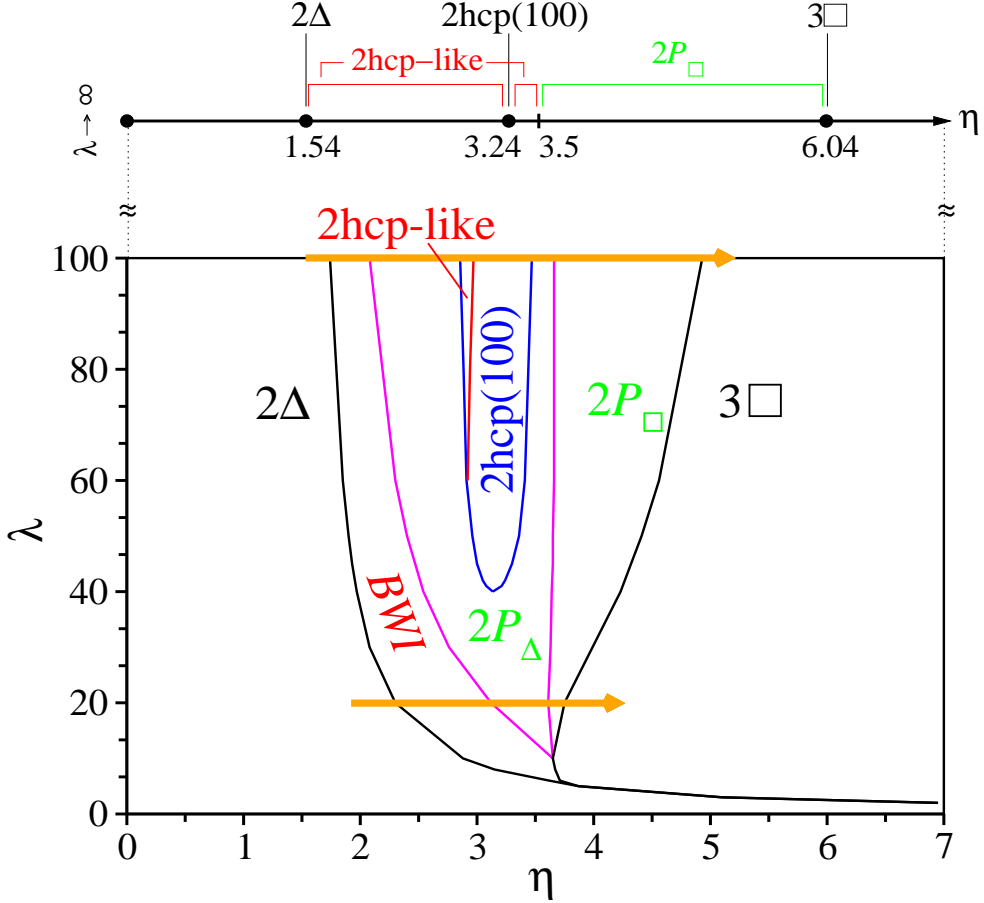}
    \caption{Zero temperature phase diagram of a Yukawa system confined between two hard plates 
      with screening strength $\lambda$ and reduced density $\eta$, in the multilayer regime between 
      2 triangular and 3 square  layers. Stability regions of the  Belgian waffle iron ({\it BWI}) phase, 
      prism phases with triangular ($2P_{\Delta}$) and square bases ($2P_{\square}$), 
      $2$hcp-like and $2$hcp(100) phases are shown. For comparison, the 
      hard-sphere limit of infinite screening is shown separately in the top. The arrow indicates a path
      of constant screening with increasing density.}
    \label{diag}
  \end{center}
\end{figure*}

Another situation concerns the confined Wigner crystal of repulsive point charges.
This classical Coulomb system has been studied when confined between two
plates and a bilayer structure is {\it always} stable \cite{footnote_1}
even for high separation 
distances $h$ \cite{Goldoni}. Here we consider a Yukawa system interpolating between 
the Coulomb interaction and hard spheres \cite{footnote_2}. 
In this case, there are two possibilities for the rhombic bilayer \cite{Messina_PRL}. Experimental 
realizations can be thought of as charged colloids \cite{Schope_2008} or dusty plasma sheets \cite{Morfill}.

In this paper we study the multilayering for a Yukawa system confined between two hard walls
\cite{Klapp} interpolating between the Coulomb and hard-sphere limit. 
For low screening, appropriate to dusty plasmas, and an external parabolic potential,
multilayering was examined by computer simulation and a simple sheet model by Totsuji and coworkers \cite{Totsuji}.
Structural details of the multilayers, however, were neglected in this shell model approach. 
Here we focus on the range of plate distances between the $2\Delta$ to $3\square$ regime and resolve the fine structure
using up to four layers as a possible candidate.
Apart from prism phases and structures derived from the hexagonal-closed-packed geometry, 
we find a stable four-layer crystal with a structure resembling a Belgian waffle iron ({\it BWI}). 
This {\it BWI}-structure which consists of two outer layers and two inner 
staggered rectangular crystals does not exist for hard spheres and unscreened charges but is stable 
in the intermediate screening regime. 
A control over the achievable crystalline
multilayered structures has important applications to the fabrication of nanosieves
and filters with desired porosity \cite{Goedel}.

In our model, we consider $N$ classical point-like particles interacting via the Yukawa pair-potential 
%
\begin{equation}
  \label{Yukawa_pot}
  V(r) = V_{0} \dfrac{e^{-\kappa r}}{\kappa r} \,, 
\end{equation}
%
where $r$ is the interparticle distance, $1/\kappa$ the screening length, 
and $V_0$ denotes an energy amplitude. 
The particles are confined between two parallel hard plates of area $A$ and separation $D$ which is taken along 
the $z$-direction for convenience. At zero temperature, for a given reduced area
density $\eta = N D^2/A$, the system will minimize its total
potential energy and the resulting optimal structure will be independent on the single 
energy scale $V_0$ but will depend on the reduced inverse screening length $\lambda = \kappa D$.
By varying this parameter  $\lambda$, one can interpolate between the unscreened Coulomb limit
($\lambda \to 0$) and the hard-sphere limit ($\lambda \to \infty$) 
where the interaction is getting very harsh.

For a given screening parameter $\lambda$ and reduced density $\eta$ we have performed lattice
sum minimizations of a set of candidates. While the stability regimes of mono- and bilayers have been addressed
in previous work \cite{Messina_PRL}, we focus here on the multilayer regime beyond the stability 
of the staggered triangular bilayer $2\Delta$ phase. This can be achieved by increasing the reduced density $\eta$. 
There is another region at even higher $D$ where the trilayer phase $3\square$ is stable. Our major goal is to explore
the corresponding distance range between the $2\Delta$ and $3\square$ phase and check for the stability
of intervening crystalline multilayers.

As possible candidates for our lattice sum minimization problem, we consider 
three dimensional crystals with a two dimensional periodicity in $x$ and $y$-direction 
whose primitive cell is a parallelogram containing $n$ particles. 
This parallelogram is spanned by the two lattice vectors ${\bf a}=a(1,0,0)$ and 
${\bf b}=a\gamma(\cos\theta,\sin\theta,0)$, 
where $\gamma$ is the aspect ratio ($\gamma = |{\bf b}|/|{\bf a}| = b/a$) and $\theta$ is the angle between 
${\bf a}$ and ${\bf b}$. 
Furthermore the $n$ particles are distributed, 
not necessarily evenly, on $m$ layers in the $xy$-plane. 
Hereby we restrict ourselves to layered situations 
with an up-down inversion symmetry in the averaged occupancy reflecting the up-down symmetry
of the confining slit. Under this sole restriction, we consider 
possible candidates with $n=2,...,8$ and $m=1,...,4$ up to symmetric four-layer 
structures with a basis of up to 8 particles. 
For given $\eta$ and $\lambda$, the total potential energy per particle is minimized with respect to the 
particle coordinates of the basis and the cell geometry ($\gamma$ and $\theta$).
The resulting stability phase diagram is shown in fig. \ref{diag} \cite{footnote_3}.

\begin{figure}
  \includegraphics[width=8.5cm]{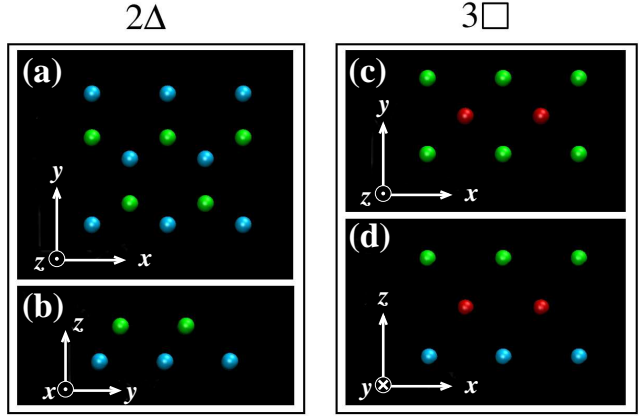} 
  \caption{(a) and (b): Top view and side view of the $2\Delta$ phase. 
      (c) and (d): Top view and side view of the $3\square$ phase. Note that the layer distance does not 
      necessarily coincide with the length of the in-plane square. Particles in different  layers are drawn
      in different colors (green, red, blue).}
  \label{2D3Q}
\end{figure}

First, the boundary of the stability domain of the staggered triangular phase $2\Delta$  and the
staggered quadratic $3\square$ are presented, see fig.\ \ref{diag}. For convenience, the structure of these two phases are 
sketched by a top and side view in fig.\ \ref{2D3Q}. In between, we observe various  stable four-layer 
structures. For low screening ($\lambda \lesssim 4$), however, there is a direct transition 
from $2\Delta$ to $3\square$ without any intermediate multilayered crystal 
consistent with earlier simulations and theoretical results
\cite{Totsuji}. The $2\Delta \to 3\square$ phase boundary moves to higher densities
as the screening $\lambda$ is lowered  
so that there are only  bilayers for the unscreened Coulomb system $(\lambda \to 0)$, see fig.\ \ref{diag} 
\cite{footnote_4}.

\begin{figure}
  \includegraphics[width=5cm]{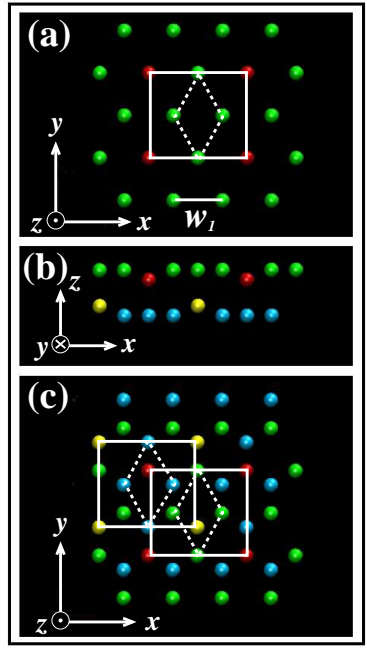} 
  \caption{Different views of the Belgian waffle iron ({\it BWI}) structure: 
    (a) Top view of an outer layer and the next inner 
    layer. The latter corresponds to a rectangular lattice (red particles) with a rhombic decoration (green particles). 
    (b) Side view of all four layers. (c) Top view of all four layers. The width $w_1$ of the rhombus is shown in (a).}
  \label{BWI}
\end{figure}

We now discuss the more complicated intermediate structures of fig.\ \ref{diag}. 
At finite $\lambda \gtrsim 4$,
the $2\Delta$ transforms discontinuously into a tetra-layered structure ($m=4$)
with $n=8$. This novel structure  shown in
fig.\ \ref{BWI} is characterized 
by two inner staggered rectangular lattices (see the red and yellow spheres in fig.\ \ref{BWI})
and two outer layers with rhombic stripes which are centered relative to the inner rectangular lattice.
In principle this structure can be obtained by a continuous transformation of the $2\Delta$ phase.
In analogy to Belgian waffles which possess an internal rectangular structure we call this
double staggered situation a  "{\it Belgian waffle iron}" (\textit{BWI}) structure, as the corresponding iron
mimics the outermost structure qualitatively \cite{footnote_5}.
The rhombus in the outermost layers is almost symmetric,
i.e. the corresponding anisotropy as characterized by the ratio of 
width $w_1$ (see fig.\ \ref{BWI}) and side length of the rhombic
is close to 1 (within 1 percent) and only slowly grows with increasing density $\eta$.

\begin{figure}
  \includegraphics[width=8.5cm]{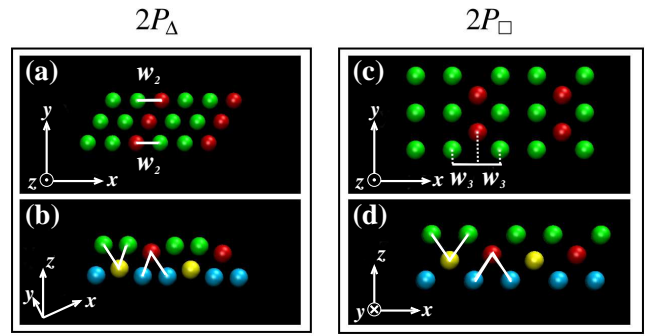} 
  \caption{Different views of the prism phases with triangular and quadratic shaped basis: 
    (a) Top view of an outer layer and the next inner layer with the appropriate width $w_2$ and 
    (b) perspective side view of $2P_{\Delta}$.
    (c) Top view of an outer layer and the next inner layer with the appropriate width $w_3$ and 
    (d) side view of $2P_{\square}$. 
    The alternating prism like arrays are indicated with white lines in (b) and (d).}
  \label{PRSM}
\end{figure}

Next, prism phases with a triangular basis shape 
$2P_{\Delta}$ and a square basis shape $2P_{\square}$ each with $n=6$ and $m=4$ (see fig.\ \ref{PRSM}), are stable.
In detail, these phases consist of alternating double layered prism-like arrays with a 
triangular or square basis structure (see fig.\ \ref{PRSM}). Within the stability regime of 
prism phases the distances $w_2$ and $w_3$ indicated in  fig.\ \ref{PRSM} change about 
few percents such that with increasing $\eta$ one can notice a slight decrease of these distances. 

\begin{figure}
  \includegraphics[width=8.5cm]{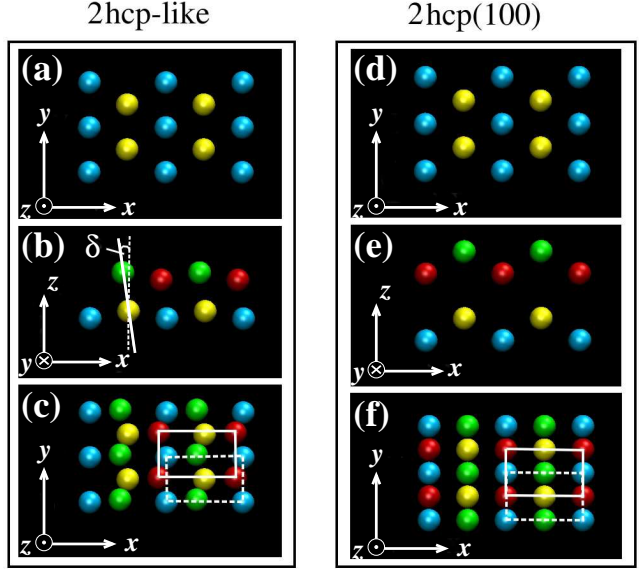} 
  \caption{Top views (a), (c) and side view (b) of $2$hcp-like as well as
    top views (d), (f) and side view (e) of $2$hcp(100). For  $\delta \to 0$, one recovers 
    $2$hcp(100). The angle $\delta$ results from the additional shift in $x$-direction of 
    the rectangular bilayers and is shown in (b).}
  \label{HCP}
\end{figure}

Finally there are tetralayered structures derivable from the hcp lattice as discussed recently 
in ref.\ \cite{Manzano_2007,Fontecha_2007} which we accordingly call $2$hcp-like and $2$hcp(100) phases (see fig.\ \ref{HCP}).
The (100)-plane of the hexagonal-closed-package, possessing a rectangular shaped basis as sketched in fig.\ 
\ref{HCP}, can be thought of as two double layered rectangular arrays which lie on the 
top of each other but shifted to each other in the $y$-direction. 
The rectangular bilayers of the $2$hcp-like structure, on the other hand, 
experience a further shift in $x$-direction, resulting in  an angle $\delta$ (see fig.\ \ref{HCP}).

For increasing screening constant $\lambda$, the topology of the phase diagram reveals the following cascades
interpolating between $2\Delta \to 3\square$:
%
%
%
%
\begin{eqnarray}
  \nonumber
  & 2\Delta \to 3\square, \\ 
  \nonumber
  & 2\Delta \to BWI \to 3\square, \\
  \label{cascade1}
  & 2\Delta \to BWI \to 2P_{\Delta} \to  2P_{\square} \to  3\square, \\
  \nonumber
  & 2\Delta \to BWI \to 2P_{\Delta} \to 2{\rm hcp}(100) \to 2P_{\Delta} \to 2P_{\square} \to  3\square
\end{eqnarray}
with a remarkable reentrance of the $2P_{\Delta}$ and finally
\begin{eqnarray}
  \nonumber
  & 2\Delta \to BWI \to 2P_{\Delta} \to 2\mbox{hcp-like} \to 2\mbox{hcp}(100) \\ 
  & \to 2P_{\Delta} \to 2P_{\square} \to  3\square.
  \label{fullcascade}
\end{eqnarray}

The hard-sphere limit involves the cascade
\begin{equation}
\nonumber
2\Delta \to 2\mbox{hcp-like} \to 2\mbox{hcp}(100) \to 2\mbox{hcp-like} \to 2P_{\square} \to 3\square
\end{equation}
which is qualitatively different from those at finite screening
as the $BWI$ and the $2P_{\Delta}$ are missing. Therefore even at high $\lambda \approx 100$
there are considerable deviations from the ultimate hard-sphere limit.

\begin{figure}
  \includegraphics[width=8.5cm]{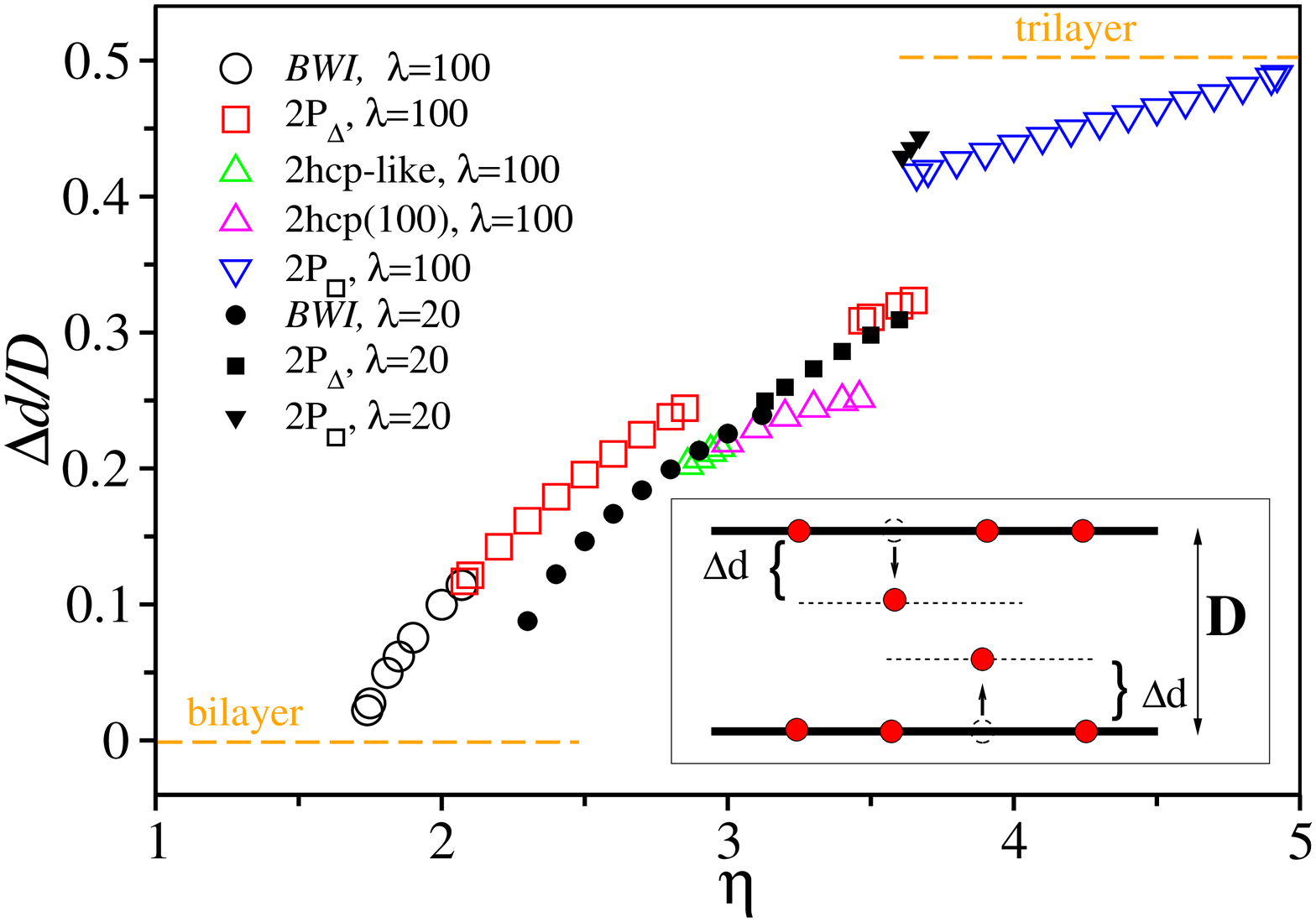} 
  \caption{Reduced distance $\Delta d/D$ between the outermost and the next inner
    layer as a function of density $\eta$ for fixed $\lambda=20$ and $\lambda=100$. The limits of a trilayer and bilayer 
    situations are indicated by the dashed lines. The geometric meaning of $\Delta d$ is shown in the 
    inset, developing of the distance between inner and outer layers.}
  \label{u_verlauf}
\end{figure}

All transitions involved here for finite $\lambda$ are first order except the $2$hcp-like $\to$ $2$hcp(100) one which is continuous.
This is illustrated in  fig.\ \ref{u_verlauf}
where the distance $\Delta d$ between the outermost and the next inner layer is shown along the arrows indicated in fig.\ 
\ref{diag}, i.e. for increasing density $\eta$ at fixed screening $\lambda=20$ and $\lambda=100$ leading to the full
transition cascade (see (\ref{fullcascade})). 
The geometric meaning  of $\Delta d$ is sketched in the inset of fig.\ \ref{u_verlauf}.
The reduced quantity $\Delta d/D$ can serve as an order parameter for multilayering in the sense, that
$\Delta d/D =0.5$ for centered trilayers and (formally) $\Delta d/D =0$ for symmetric bilayers.
Multilayers beyond $m=3$ are characterized by an intermediate value of $\Delta d/D$. Moreover
for an equilayer-spacing, $\Delta d/D$ is $1/(m-1)$. For $\lambda=100$, clearly, the transition from $2\Delta$ to 
$BWI$ is accompanied with a jump from zero to $\Delta d/D =0.0221$ ($\Delta d/D = 0.0933$ for $\lambda=20$) implying a first-order transition 
although a second-order transition would have
been  not precluded by lattice symmetry. The subsequent $BWI \to 2P_{\Delta}$ transition is discontinuous
(though the jump in $\Delta d/D$ is tiny in fig.\ \ref{u_verlauf}) while the $2$hcp-like $\to$ $2$hcp(100)
transition is continuous. The latter transition occurs as the angle $\delta$ approaches 
zero as indicated in fig.\ \ref{HCP}. 
For $\lambda=20$, there are only discontinuous transitions along the cascade (\ref{cascade1}).
In the limit of hard spheres, volume fraction calculations  among the considered candidates show that the transitions 
$2\Delta \to$ $2$hcp-like $\to$ $2$hcp(100) $\to$ $2$hcp-like are of second order, while the transition 
$2$hcp-like $\to$ $2P_{\square}$ is of first order. The transition $2P_{\square} \to 3\square$ occurs in the 
hard-sphere limit continuously, as opposed to finite $\lambda$.

To summarize, we investigated the stability phase diagram of a confined classical one-component Yukawa system in the 
transition regime $2\Delta \to 3\square$. Various crystalline tetra-layer structures are intervening
including a novel one with a "Belgian waffle iron structure" ($BWI$), phases of alternating prisms
$2P_{\Delta}$, $2P_{\square}$, and structures derived from a bulk hcp structure ($2$hcp-like and $2$hcp(100)). 
For the unscreened Coulomb system only bilayers are stable while in the opposite (hard-sphere) limit 
the sequence for increasing density is: $2\Delta \to$ $2$hcp-like $\to$ $2$hcp(100) $\to$ $2$hcp-like 
$\to 2P_{\square} \to 3\square$ \cite{Manzano_2007}. 
Our theoretical results are verifiable in experiments on colloidal suspensions.
In fact, all phases were found in recent experimental investigations except the \textit{BWI} structure which, however, 
resembles experimentally determined superstructures \cite{Schope_2008}. Furthermore strongly interacting dust
particles in plasmas are typically in the intermediate screening regime at low reduced temperatures
and are therefore other candidates for an experimental system where the multilayer phases proposed here
are in principle observable.

Future work should focus on the inclusion of higher layer numbers $m$.
We have examined the special case with $m=6$ layers which was called pre-3$\square$ in ref.\ \cite{Manzano_2007}. 
This structure, however, was not found to be stable in the range of $\lambda$ considered in this work.
Next, charged walls will lead to a soft exponentially-screened wall-particle 
interactions \cite{wedge} rather than a hard wall
which would introduce more parameters 
in the corresponding model \cite{Totsuji}. In this situation, the multilayering scenario is expected to favor phases
which have more weight in the inner layers. The effect of finite temperature 
\cite{Messina_PRE_2006} needs more 
investigation; temperature can be viewed as an additional coordinate in the two-dimensional phase diagram presented in 
fig.\ \ref{diag}. We expect, however, no drastic change in the phase diagram topology. Finally we remark
that genetic minimization algorithms \cite{Gottwald} might be a versatile tool to explore even more complicated 
crystalline layer structures.

We thank T. Palberg, H. J. Sch\"ope, A. Wynveen and F. Ramiro-Manzano for helpful discussions.
This work was supported by the DFG (SFB TR6, project D1).


\begin{thebibliography}{100}


\bibitem{Haase} S. Dietrich and A. Haase,   
  Phys. Reports {\bf 260}, 1 (1995).

\bibitem{Klafter} J. M. Drake and J. Klafter,  
  Phys. Today {\bf 43}, 46 (1990).

\bibitem{Zangi} R. Zangi and S. Rice,
  Phys. Rev. E {\bf 61}, 660 (2000).

\bibitem{Ayappa} K. G. Ayappa and C. Ghatak, 
  J. Chem. Phys. {\bf 117}, 5373 (2002).

\bibitem{Cornelissens} Y. G. Cornelissens, B. Partoens and F. M. Peeters, 
  Physica E {\bf 8}, 314 (2000).

\bibitem{Schmidt1} M. Schmidt and H. L{\"o}wen,
  Phys. Rev. Lett. {\bf 76}, 4552 (1996).
  
\bibitem{Nelson} T. Chou and D. R. Nelson, 
  Phys. Rev. E {\bf 48}, 4611 (1993).
  
\bibitem{Frey} L. Radzihovsky, E. Frey and D. R. Nelson, 
  Phys. Rev. E {\bf 63}, 031503 (2001).

\bibitem{Pieranski} P. Pieranski, L. Strzlecki and B. Pansu,
  Phys. Rev. Lett. {\bf 50}, 900 (1983).

\bibitem{Schmidt2} M. Schmidt and H. L{\"o}wen,
  Phys. Rev. E {\bf 55}, 7228 (1997).

\bibitem{Neser} S. Neser, C. Bechinger, P. Leiderer and T. Palberg, 
  Phys. Rev. Lett. {\bf 79}, 2348 (1997).

\bibitem{Fortini} A. Fortini and M. Dijkstra, 
  J. Phys.: Condens. Matter {\bf 18}, L371 (2006).

\bibitem{Manzano_2007} F. Ramiro-Manzano, E. Bonet, I. Rodriguez and F. Meseguer, 
  Phys. Rev. E {\bf 76} 050401 (2007).

\bibitem{Fontecha_2007} A. Barreira Fontecha, T. Palberg and H. J. Sch{\"o}pe, 
  Phys. Rev. E {\bf 76}, 050402 (2007).

\bibitem{Manzano_2006} F. Ramiro-Manzano, F. Meseguer, E. Bonet and I. Rodriguez,
  Phys. Rev. Lett. {\bf 97},  028304 (2006).

\bibitem{Schope_2006} H. J. Sch\"ope, A. Barreira Fontecha, H. K\"onig, J. Marques Hueso and R. Biehl,
  Langmuir {\bf 22}, 1828 (2006).

\bibitem{footnote_1} There is a simple and clear electrostatic argument to explain the exclusive stability
  of bilayers for charges confined between (charged or uncharged) hard walls. 
  Note that two equally charged walls do {\it not} generate any electric field 
  within the slit, and consequently do not alter the stable structure obtained at any other 
  surface charge (including neutral walls). Hence, if one considers the special case of two walls 
  corresponding to neutralizing backgrounds, then the ground state structure is always a bilayer. 

\bibitem{Goldoni} G. Goldoni and F. M. Peeters, 
  Phys. Rev. B {\bf 53}, 4591 (1995).

\bibitem{footnote_2} Note that in our model we have point-like particles. 
  Nevertheless, at infinite screening ($\lambda \to \infty$), by taking an effective hard-core diameter 
  corresponding to the smallest lattice constant, one expects to recover the phase behavior of hard spheres.

\bibitem{Messina_PRL} R. Messina and H. L{\"o}wen, 
  Phys. Rev. Lett. {\bf 91}, 146101 (2003).

\bibitem{Schope_2008} A. Barreira Fontecha and H. J. Sch\"ope,
  Phys. Rev. E {\bf 77}, 061401 (2008).

\bibitem{Morfill} G. E. Morfill, A. V. Ivlev, M. Rubin-Zuzic, C. A. Knapek, R. Pompl, T. Antonova and H. M. Thomas,  
  Appl. Phys. B (Lasers and Optics) {\bf 89}, 527 (2007).

\bibitem{Klapp} For a recent study of the fluid phase, see: 
  S. H. L. Klapp, Y. Zeng, D. Qu and R. von Klitzing,
  Phys. Rev. Lett. {\bf 100}, 118303 (2008).
 
\bibitem{Totsuji} H. Totsuji, T. Kishimoto and C. Totsuji, 
  Phys. Rev. Lett. {\bf 78}, 3113 (1996).

\bibitem{Goedel} P. Tierno, K. Thonke and W. A. Goedel,
  Langmuir {\bf 21}, 9476 (2005).

\bibitem{footnote_3} One can calculate density jumps \cite{Graf1998,vanRoij1999,Warren2000} 
  between the different phases. These turn out to be very small except close to the lower ends of phase stability, 
  e.g.\ at $\lambda=10$, $\eta=3.6$ for the $2P_{\square}$, and are not shown in figure 1 for clarity.

\bibitem{Graf1998} H. Graf and H. L\"owen,
  Phys. Rev. E {\bf57}, 5744 (1998).

\bibitem{vanRoij1999} R. Van Roij, M. Dijkstra and J. P. Hansen, 
  Phys. Rev. E {\bf 59}, 2010 (1999).

\bibitem{Warren2000} P. B. Warren, J. Chem. Phys. {\bf 112}, 4683 (2000).

\bibitem{footnote_4} As a side remark, we found that a triangular (projected) bilayer, 
  proposed as a stable structure for very small $\eta$ in \cite{Goldoni}, is always 
  energetically beaten by a buckled bilayered phase.

\bibitem{footnote_5} We remark that the BWI structure can also be thought of arising 
  from a buckled layer adjacent to a structured layer.

\bibitem{wedge} H. L{\"o}wen, A. H\"artel, A. Barreira Fontecha, 
  H. J. Sch\"ope, E. Allahyarov and T. Palberg,
  J. Phys.: Condens. Matter  {\bf 20}, 404221 (2008).

\bibitem{Messina_PRE_2006} R. Messina and H. L\"owen,
  Phys. Rev. E {\bf 73}, 011405 (2006).

\bibitem{Gottwald} D. Gottwald, C. N. Likos, G. Kahl and H. L{\"o}wen,
  J. Chem. Phys. {\bf 122}, 074903 (2005).

\end{thebibliography}
\end{document}